\documentclass[aps,twocolumn,noshowpacs,noshowkeys,letterpaper,superscriptaddress]{revtex4}%
\usepackage{amsfonts}
\usepackage{amsmath}
\usepackage{amssymb}
\usepackage{graphicx}
\usepackage{textcomp}%
\begin{document}
\title{Cooling of a micro-mechanical oscillator using radiation pressure
induced dynamical back-action}

\author{A.\ Schliesser}
\affiliation{Max Planck Institut f\"ur Quantenoptik, 85748 Garching,
Germany}
\author{N.\ Nooshi}
\affiliation{Max Planck Institut f\"ur Quantenoptik, 85748 Garching,
Germany}
\author{P.\ Del'Haye}
\affiliation{Max Planck Institut f\"ur Quantenoptik, 85748 Garching,
Germany} %
\author{K.J.\ Vahala}%
\email{vahala@caltech.edu}%
\affiliation{Department of Applied Physics, California Institute of
Technology, Pasadena, CA 91125 USA}%
\author{T.J.\ Kippenberg}
\email{tjk@mpq.mpg.de} %
\homepage{www.mpq.mpg.de/k-lab/}%
\affiliation{Max Planck Institut f\"ur Quantenoptik, 85748 Garching, Germany} %


\begin{abstract}
Cooling of a 58 MHz micro-mechanical resonator from room temperature to 11 K
is demonstrated using cavity enhanced radiation pressure. Detuned pumping of
an optical resonance allows enhancement of the blue shifted motional sideband
(caused by the oscillator's Brownian motion) with respect to the red-shifted
sideband leading to cooling of the mechanical oscillator mode.\ The reported
cooling mechanism is a manifestation of the effect of radiation pressure
induced dynamical backaction. These results constitute an important step
towards achieving ground state cooling of a mechanical oscillator.

\end{abstract}
\maketitle
\revised{}

Cooling of micro-mechanical resonators such as cantilevers is an
important prerequisite for studies ranging from highly sensitive
measurements of forces \cite{Rugar2004} and displacement
\cite{Cleland1998} to observing quantum mechanical phenomena of
macroscopic objects \cite{Marshall2003} and gravitational wave
detection. Early work recognized the possibility to cool a
mechanical resonator mode by radiation pressure, via use of an
active feedback scheme \cite{Cohadon1999,Mancini1998}, in close
analogy to stochastic cooling of charged particles. However,
techniques which cool a mechanical oscillator intrinsically via
radiation pressure, as routinely achieved in atomic laser cooling---%
while proposed theoretically
\cite{Braginsky1977,Kippenberg2005,Braginsky2002}---have never been
demonstrated. Here we report such a technique based on radiation
pressure \cite{Braginsky2002} and apply it to cool a 58-MHz
micro-mechanical oscillator in the form of a toroid cavity from room
temperature to 11 K.

The present work falls within the setting of high finesse
opto-mechanical systems, which couple a mechanical oscillator (such
as a cantilever \cite{Craighead2000,Huang2003,Metzger2004}, an
internal mode of a mirror \cite{Cohadon1999} or a microcavity
\cite{Rokhsari2005}) to an optical cavity field by means of a high
finesse optical resonator as shown schematically in Fig.\ 1(B). The
motion of the mirror renders the cavity transmission position
dependent. Indeed, the use of a Fabry $\mathrm{P\acute {e}rot}$
interferometer is among the most sensitive methods for displacement
measurements \cite{Braginsky1977} and the underlying principle of
the laser interferometer gravitational wave observatory, LIGO. This
monitoring however can become unstable owing to radiation pressure,
which can lead to regenerative oscillations of the mechanical
eigenmodes as first predicted by Braginsky \cite{Braginsky2001}.
This phenomenon is a manifestation of the effect of
\textit{dynamical back-action} \cite{Braginsky1977,Braginsky1992}.
In fact, while effects of radiation pressure related phenomena (such
as bistability \cite{Dorsel1983}) have been observed for more than
two decades---and have been subject of various theoretical studies
\cite{Mancini2002,Mancini1998,Marshall2003}---the dynamical
back-action caused by radiation pressure has only recently been
observed in toroid micro-cavities on a chip
\cite{Rokhsari2005,Carmon2005,Kippenberg2005}. These toroidal
microresonators \cite{Vahala2003,Armani2003} (cf.\ Fig.\ 1) possess
ultra-high-Q whispering gallery type optical modes while
simultaneously exhibiting micro-mechanical resonances in the
radio-frequency domain. Because of the curved nature of the
dielectric cavity, the circulating photons exert a radial force on
the cavity sidewalls, thereby coupling the optical and mechanical
degree of freedom, analogous to a Fabry $\mathrm{P\acute {e}rot}$
cavity with a movable mirror (Fig.\ 1(B)). Owing to the resonator's
optical photon lifetime being similar to the mechanical oscillation
period, these devices operate in a regime where cavity retardation
effects cannot be neglected, which has enabled the study of
radiation pressure induced oscillations
\cite{Rokhsari2005,Kippenberg2005,Carmon2005} and related effects
\cite{Marquardt2006,Braginsky1977} in an experimental setting. While
dynamical back-action can lead to mechanical oscillations
\cite{Rokhsari2005} it can also as shown here lead to cooling of the
mechanical mode as predicted in earlier work
\cite{Braginsky2002,Kippenberg2005}.

\begin{figure}[hbp]
\begin{center}
\includegraphics[width=.7\linewidth]{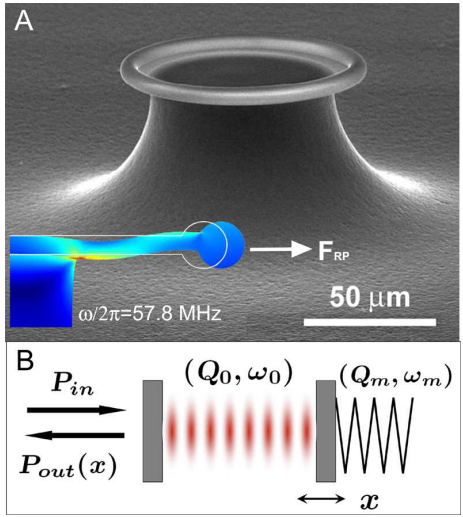}
\end{center}
\caption{(color online)\ Main panel: A scanning electron micrograph
of a toroid microcavity on a chip. Inset: Finite element simulation
showing the stress (color coded) and strain of the 57.8-MHz
breathing mode of the device. The strain is exaggerated for clarity.
Note that radiation pressure exerts a radial force owing to total
internal reflection of the confined light. (B) A
Fabry-$\mathrm{P\acute{e}rot}$ equivalent diagram of the experiments.}%
\end{figure}

The mechanical mode under consideration in this study is a radial
breathing mode with resonance frequency ($f_{m}=\omega_{m}/2\pi$) of
57.8 MHz. Since the mechanical mode is in thermal equilibrium with
the environment (T =300 K), it follows from the classical
equipartition theorem that the mechanical oscillator undergoes
Brownian motion with a root mean square displacement of $\langle
x^{2}\rangle^{1/2}=\sqrt{\frac{k_{B}T}{m\omega_{m}^{2}}}$, where
$k_{B}$ is the Boltzmann constant. For the present micro-resonator,
whose effective mass is $m_\mathrm{eff}=1.5\times10^{-11}$kg (as
determined independently by finite-element simulation) this
amplitude corresponds to $\sim 5\times10^{-14}$m. Despite the small
amplitude the thermally excited oscillations are readily observable
in the transmitted light, when a cavity mode is excited using a
tapered optical fiber. Operating the cavity detuned, the amplitude
fluctuations of the mechanical resonator cause a change in the
cavity transmission that enables extraction of the mechanical
resonator characteristics (resonance frequency $\omega_{m}/2\pi$,
line-width $\gamma/2 \pi=\omega_{m}/2\pi Q_{m}$, and displacement
spectral density $\langle x_{\widetilde{\omega}}^{2}\rangle$).
Figure 2 shows displacement monitoring of the micro-resonator
obtained for varying power levels at a constant red-detuning of the
laser frequency $\omega/2\pi$, with respect to the cavity resonance,
$\omega_{0}/2\pi$. The normalized detuning $\Delta\omega
\tau=(\omega-\omega_{0})\tau$, where $\tau$ is the photon lifetime,
is ca. $-0.5$ for these measurements. From fits to the noise spectra
at low pump power (10 $\mathrm{\mu W}$), the resonance frequency of
$57.8$ MHz and an intrinsic mechanical quality factor ($Q_{m}$) of
2890 were extracted. The optical line-width at critical coupling was
$\sim$ $50$ MHz, equivalent to an optical quality factor of
$Q=\omega_{0}\tau=6.2\times10^{6}$ . Strikingly, when increasing the
power of the red-detuned laser, a strong decrease in thermal
displacement was observed (Fig.\ 2, main panel). Concomitant a
strong increase in the mechanical line-width (damping) was also
observed (cf.\ Fig.\ 2(B)), which in the present case can be used to
infer an \emph{upper limit} on the effective oscillator temperature
\cite{Cohadon1999}. Indeed, we confirmed that the area underneath
the Lorentzian noise spectrum reduces by at least a factor
$\times\frac{\gamma}{\gamma_\mathrm{eff}}$ (where
$\gamma_\mathrm{eff}$ is the effective linewidth \cite{Cohadon1999})
which is a further measure of the effective oscillator temperature
(since $T_\mathrm{eff}k_\mathrm{B}=\int m_\mathrm{eff}\langle
x_{\widetilde
{\omega}}^{2}\rangle\widetilde{\omega}^{2}d\widetilde{\omega}$ ).
For the highest pump power ($\sim$2 mW at 970 nm), the effective
temperature was reduced from 300 K to 11 K.

\begin{figure}[tbp]
\begin{center}
\includegraphics[width=\linewidth]{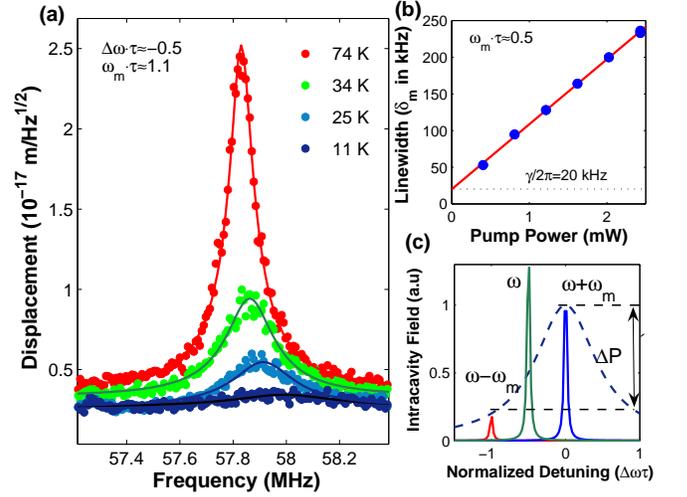}
\end{center}
\caption{(color online) Main figure shows the normalized, measured
noise spectra around the mechanical resonance frequency for
$\Delta\omega\tau\approx-0.5$ and varying power (0.25, 0.75, 1.25
and 1.75 mW). The effective temperatures were inferred using
mechanical damping, with the lowest attained temperature being 11 K.
Panel (B): Increase in the line-width (damping) of the 57.8-MHz-mode
as a function of launched power, exhibiting the expected linear
behaviour (cf.\ Eq.\ (1)). Panel (C): Physical origin of the
observed cooling
mechanism due to the asymmetry in the motional sidebands.}%
\end{figure}

To substantiate our claim that indeed radiation pressure is
responsible for the observed cooling, a theoretical coupled mode
model based on radiation pressure is used \cite{Kippenberg2005} and
compared to our findings. The oscillating cavity (with frequency
$\omega_{m}$ ) produces both Doppler up-shifted
($\omega_{AS}=\omega+\omega_{m}$ ) anti-Stokes photons and
down-shifted ( $\omega_{S}=\omega-\omega_{m}$) Stokes photons
(equivalent to the motional sidebands of trapped ions). If the
cavity is pumped red detuned (i.e. $\omega<\omega_{0}$ ), the cavity
resonance will suppress scattering into the red sideband, while
enhance scattering into the blue sideband. This situation is
depicted schematically in Fig.\ 2(C). This asymmetry in the
sidebands leads to a \emph{net transfer} of power from the
mechanical oscillator to the light field and causes \emph{cooling}
of the mechanical resonator mode. The cooling power can be derived
by noting that the generated Stokes and anti-Stokes sidebands
produce a time varying radiation pressure force, whose quadrature
component upon red-detuning can extract power
($\widetilde{P}=\langle\frac{dx}{dt}F_{RP}\rangle$ ) from the
mechanical oscillator mode. The corresponding cooling rate
($\Gamma=-\frac{\widetilde{P}}{\langle E\rangle})$ is given by%
\begin{equation}
\Gamma=\frac{-\mathit{F}^{2}8n^{2}\omega_{0}}{m_\mathrm{eff}c^{2}\omega_{m}}C\left(
\frac{1}{4\tau^{2}\Delta\omega_{s}^{2}+1}-\frac{1}{4\tau^{2}\Delta\omega
_{as}^{2}+1}\right)  P_{\text{in}}.%
\end{equation}
Here $P_{\text{in}}$ denotes the launched power into the fiber,
$\mathit{F}$ denotes finesse, and $1/\tau_{ex}$ is the rate of
coupling into the cavity from the fiber and $\Delta\omega$ the laser
detuning. Furthermore the coupling parameter $C$ is introduced,
$C\equiv\frac{\tau/\tau_{\mathrm{ex}}}{4\tau
^{2}\Delta\omega^{2}+1}\in\lbrack0..1]$ and $m_\mathrm{eff}$ is the
effective mass of the mechanical mode under consideration. The
detuning of the Stokes and anti-Stokes photons are given by
$\Delta\omega_{S}=\Delta\omega-\omega_{m}$ and
$\Delta\omega_{AS}=\Delta\omega+\omega_{m}$. We emphasize that
efficient cooling is characterized by $\omega_{m}\tau\approx1$ . It
is interesting to note that the physical process which gives rise to
cooling is analogous to cooling of atoms or molecules via coherent
scattering \cite{Vuletic2000,Maunz2004}. The final temperature of
the mechanical oscillator in the presence of cooling is given by the
balance of heating by the reservoir ($\langle
P\rangle=\omega_{m}\frac{k_{B}T_{R}}{Q_{m}}$, as described by the
fluctuation dissipation theorem) and the radiation pressure
induced cooling rate \cite{Cohadon1999}:%
\begin{equation}
\frac{T_\mathrm{eff}}{T_{R}}\approx\frac{\gamma}{\gamma+\Gamma}.
\end{equation}
The presence of cooling thus changes the observed line-width of the
mechanical resonances ($\gamma_\mathrm{eff}=\gamma+\Gamma$) and
causes a reduction in the Brownian amplitude. In addition, the in
phase component of radiation pressure gives rise to a change in
mechanical
resonance frequency ($\omega_\mathrm{eff}=\omega_{m}+\Delta\omega_{m}$ ):%
\begin{equation}
\Delta\omega_{m}=\frac{8n^{2}\mathit{F}^{2}\omega_{0}}{m_\mathrm{eff}c^{2}\omega_{m}%
}\tau C\left(  \frac{\Delta\omega_{s}}{4\tau^{2}\Delta\omega_{s}^{2}+1}%
+\frac{\Delta\omega_{as}}{4\tau^{2}\Delta\omega_{as}^{2}+1}\right)
P_{\text{in}}%
\end{equation}
To verify these additional theoretical predictions a series of
experiments were performed. First, the dependence of the cooling
rate $\Gamma$ on the detuning was verified (Eq.\ (1) and Fig.\ 3).
For negative detuning a clear increase in the mechanical line-width
is observed (i.e.\ cooling) and for positive detuning a decrease in
line-width. Furthermore, for blue detuning, regions exists in which
the radiation pressure induced amplification rate exceeds the
intrinsic mechanical dissipation rate $\gamma$, leading to the
regime of parametric oscillation instability
\cite{Rokhsari2005,Kippenberg2005,Carmon2005}. Using Eq.\ (1) and
leaving only the effective mass a free parameter (which agreed well
with the simulated value), accurate agreement with the
experimentally observed change in mechanical line-width (damping)
was obtained. This represents the first experimental evidence that
indeed radiation pressure is responsible for the observed cooling.

\begin{figure}[htbp]
\begin{center}
\includegraphics[width=.85\linewidth]{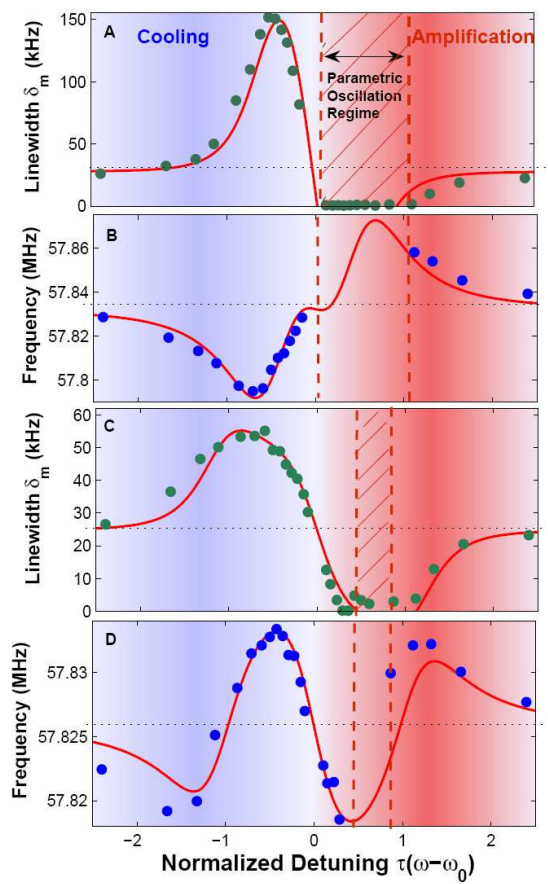}
\end{center}
\caption{(color online) Mechanical linewidth (A) and mechanical
resonance frequency (B) of the $57.8$-MHz radial breathing mode as a
function of (normalized) detuning when the laser is tuned over a 113
MHz wide optical cavity resonance (corresponding to
$\omega_{m}\tau=0.5$). The optical power was $P=1.0\,\mathrm{mW}$.
The dashed region denotes occurrence of the parametric oscillation
instability. For \mbox{$\Delta\omega < 0$} the 58-MHz resonance is
being cooled. Plot (C,D) show the characteristics
($\omega_\mathrm{eff}/2\pi$ and linewidth) of the same mechanical
mode, but for scanning over a narrower (50 MHz linewidth) optical
resonance (corresponding to $\omega_{m}\tau \approx1.1)$ with
$P=0.13\,\mathrm{mW}$. Solid lines are theoretical predictions based
on Eqs.\ (1)-(3). The only free parameter was the effective mass,
which agreed well with the
simulated value.}%
\end{figure}

\begin{figure}[tbp]
\begin{center}
\includegraphics[width=\linewidth]{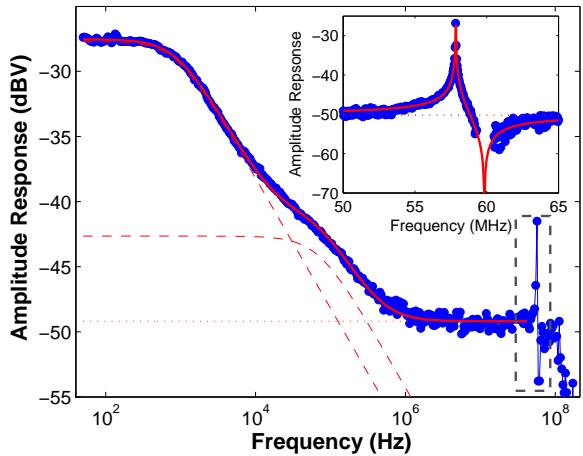}
\end{center}
\caption{(color online) Main panel: The frequency response from
0-200 MHz. The plateau occurring between 1-200 MHz is due to the
(instantaneous) Kerr nonlinearity of $\mathrm{SiO_{2}}$ (dotted
line). The cut-off at 200 MHz is due to both detector and cavity
bandwidth. Inset: magnification of main panel in vicinity of
radiation pressure response which shows the interference of the Kerr
nonlinearity and the radiation pressure driven micro-mechanical
resonator (which, on resonance, is \mbox{$\mathrm{\pi/2}$} out of
phase with the modulating pump and the instantaneous Kerr
nonlinearity). From the fits (solid lines) it can be inferred that
the radiation pressure response is ca.\
\mbox{$\mathrm{\times260}$} larger than the Kerr response.}%
\end{figure}
Next, the modifications to the rigidity of the oscillator were
measured. For $1/\tau>2\omega_{m}$, Eq.\ (3) predicts a decrease
(increase) of the mechanical frequency in the cooling
(amplification) regime. However, if the cavity bandwidth satisfies
$1/\tau<2\omega_{m}$, an unexpected behaviour is predicted; the
oscillator frequency will \emph{increase} \emph{when cooled} for
small detuning, while for the remaining detuning it is shifted to
lower frequencies. These predictions are in excellent agreement with
the reported experimental results as shown in Fig.\ 3. Keeping the
same sample but using a different optical resonance with a
line-width of 113 MHz (57.8 MHz mechanical resonance), the
transition to a pure negative and positive frequency shift in the
cooling and amplification regimes was observed as shown in Fig.\
3(B), confirming the validity of our theoretical model.

While prior work has already established that radiation pressure is
the dominant mechanism that drives mechanical oscillations in toroid
microcavities \cite{Carmon2005,Kippenberg2005,Rokhsari2005},
measurements were also carried out to assess the contribution of
thermal effects \cite{Metzger2004,Zalautdinov2001}. Thermal effects
can cause regenerative mechanical oscillation and even cooling
\cite{Metzger2004}. In order to quantify the thermal contribution at
57.8 MHz, the response of the cavity to a modulated pump in the
frequency range of 0-200 MHz was recorded (Fig.\ 4). This was
achieved by recording the shift in optical resonance due to a
modulated laser using a pump-probe scheme allowing to differentiate
different path-length-changing contributions due to the thermal
nonlinearities, radiation pressure and Kerr-effect based on their
distinct frequency response. The poles in the response at 1.6 kHz
and 119 kHz agree well with convective heat exchange to the N$_{2}$
environment surrounding the resonator and thermal conduction to
silica. For frequencies beyond 1 MHz, a plateau appears which is due
to the (instantaneous) Kerr nonlinearity as observed in prior work
\cite{Rokhsari2005a}. At 58 MHz the radiation pressure driven
mechanical response is observable, which (inset Fig.\ 4) is a factor
of $\times$ 260 stronger than the Kerr nonlinearity in the present
device. It is emphasized that this ratio is in quantitative
agreement with the theoretically \emph{predicted} Kerr-to-radiation
pressure ratio. From the well-identified frequency response of the
aforementioned thermal effects we can furthermore conclude that
their contribution to the interaction of the cavity field with the
57.8-MHz radial breathing mode is at least two orders of magnitude
too weak to explain the observed effects. Consequently, the response
measurements give unambiguous proof that such thermal effects
contribute less than 1 part in 100 to the observed cooling (or
heating) rate.

In summary we have reported radiation pressure induced cavity
cooling of a radio frequency (57.8-MHz) micro-mechanical oscillator.
Given the recent progress in high-finesse opto-mechanical systems it
is reasonable to assume that this phenomenon will become observable
in a wide variety of systems, and might provide a route to achieve
ground state cooling of a micro-mechanical oscillator.

\textit{Acknowledgements: }This work was funded via a Max Planck
Independent Junior Research Group grant, a EU\ Marie Curie Grant
(CMIRG-CT-2005-031141) and the DFG\ (NIM Initiative). The authors
gratefully acknowledge J. Kotthaus for clean-room access and J.
Alnis for technical assistance. T.\ W.\ H\"{a}nsch, K.\ Karrai and
W.\ Zwerger are thanked for stimulating discussions. \textit{Note
added:} After submission of this work, radiation pressure cooling of
a micro-mechanical mirror by dynamical back action \cite{Pinard2006,
Zeilinger2006} and active feedback \cite{Bouwmeester2006} was
reported.

\end{document}